\documentclass[english,12pt]{article}
\usepackage{physics}
\usepackage{array}
 \usepackage[normalem]{ulem}
\usepackage{graphicx}
\usepackage{amssymb}
\usepackage[english]{babel}
\usepackage{amsmath}
\usepackage{braket}
\usepackage{multirow}
\usepackage{prettyref}
\usepackage{babel}
\usepackage{units}
\usepackage[latin1]{inputenc}
\usepackage{amsfonts}
\usepackage{amssymb}
\usepackage{slashed}
\usepackage{babel}
\usepackage{color}
\usepackage[dvipsnames]{xcolor} 
\usepackage{cite}
 \def\eq#1\en{\begin{equation}#1\end{equation}}
\def\s[#1,#2]{[#1\stackrel{\star}{,}#2]}
\def\sx[#1,#2]{[#1\stackrel{\star_{x}}{,}#2]}

\newcommand{\cl}{\mathrm{cl}}
\newcommand{\gh}{\mathrm{gh}}
\newcommand{\TT}{\mathrm{TT}}
\newcommand{\BRST}{\mathrm{BRST}}

\usepackage[english]{babel}

\usepackage{setspace}

\usepackage{graphicx} 
\usepackage{graphics} 
\usepackage{float} 
\usepackage{subcaption} 

\usepackage{amsmath}
\usepackage{amssymb}
\usepackage{mathtools}
\usepackage{enumerate}
\usepackage{enumitem}
\usepackage[colorlinks=true, allcolors=black]{hyperref}

\usepackage{tensor}

\def\gsim{\mathrel{\rlap{\lower4pt\hbox{\hskip1pt$\sim$}}
		\raise1pt\hbox{$>$}}}       

\newcommand{\nc}{\newcommand}
\nc{\beq}{\begin{equation}}
\nc{\eeq}{\end{equation}}
\nc{\beqa}{\begin{eqnarray}}
\nc{\eeqa}{\end{eqnarray}}

\def\bc{\begin{center}}
\def\ec{\end{center}}

\def\to{\rightarrow}

\def\gsim{\mathrel{\mathpalette\atversim>}}

\def\bc{\begin{center}}
\def\ec{\end{center}}

\def\gsim{\mathrel{\rlap{\lower4pt\hbox{\hskip1pt$\sim$}}

    \raise1pt\hbox{$>$}}}       

\def\gsim{\mathrel{\rlap{\lower4pt\hbox{\hskip1pt$\sim$}}
    \raise1pt\hbox{$>$}}}       

\begin{document}
\makeatletter
\def\fmslash{\@ifnextchar[{\fmsl@sh}{\fmsl@sh[0mu]}}
\def\fmsl@sh[#1]#2{%
  \mathchoice
    {\@fmsl@sh\displaystyle{#1}{#2}}%
    {\@fmsl@sh\textstyle{#1}{#2}}%
    {\@fmsl@sh\scriptstyle{#1}{#2}}%
    {\@fmsl@sh\scriptscriptstyle{#1}{#2}}}
\def\@fmsl@sh#1#2#3{\m@th\ooalign{$\hfil#1\mkern#2/\hfil$\crcr$#1#3$}}

\makeatother

\hypersetup{pageanchor=false}
\thispagestyle{empty}
\begin{titlepage}
\boldmath
\begin{center}
  \Large {\bf {Quantum State of the Asymptotic Gravity Field from BRST Quantization}}
    \end{center}
\unboldmath
\vspace{0.2cm}
\begin{center}
{\large Xavier~Calmet}\footnote{E-mail: xcalmet@gmail.com}$^a$
{\large and}  {\large  Stephen D. H. Hsu}\footnote{E-mail: hsusteve@gmail.com}$^b$
 \end{center}
\begin{center}
$^a${\sl Department of Physics and Astronomy, \\
University of Sussex, Brighton, BN1 9QH, United Kingdom
}\\
$^b${\sl Department of Physics and Astronomy\\ Michigan State University, East Lansing, Michigan 48823, USA}\\
\end{center}
\vspace{5cm}
\begin{abstract}
\noindent
We derive the quantum state of the asymptotic linearized gravity field of a compact energy source, using BRST quantization. Barnich has previously derived the quantum state of the U(1) field in QED using the Gauss law constraint. The role of Gauss's law is now played by the linearized Hamiltonian and momentum constraints in gravity. For a static mass source, only the scalar Hamiltonian constraint is shifted. The quantum state corresponding to the static Newton field originates in the non-radiative scalar constraint sector, not in the propagating transverse-traceless graviton sector. We discuss implications for holography and for black hole information: 1. the exact quantum state of the compact source is encoded in the asymptotic quantum state of its gravity field 2. Hawking amplitudes are dependent, via the gravity state outside the horizon, on the black hole internal state.
\end{abstract}  
\end{titlepage}
\hypersetup{pageanchor=true}



\newpage

\section{Introduction}

The aim of the paper is to derive the quantum state of the asymptotic linearized gravity field of a compact energy source, using BRST quantization. Barnich \cite{Barnich:2010bu} (see \cite{Muck:2013orm} as well)  has previously derived the quantum state of the U(1) field in QED using the Gauss law constraint. We show that the role of Gauss's law is now played by the linearized Hamiltonian and momentum constraints in gravity. For a static mass source, only the scalar Hamiltonian constraint is shifted. The quantum state corresponding to the static Newton field originates in the non-radiative scalar constraint sector, not in the propagating transverse-traceless graviton sector. We discuss implications for holography and for black hole information: 1. the exact quantum state of the compact source is encoded in the asymptotic quantum state of its gravity field 2. Hawking amplitudes are dependent, via the gravity state outside the horizon, on the black hole internal state. This work is a non-perturbative (albeit still weak-field) generalization of work deriving quantum hair (i.e., corrections to the potential or metric) using perturbation theory \cite{Calmet:2021stu,Calmet:2021cip,Calmet:2022swf,Calmet:2022bpo,Calmet:2023gbw,Calmet:2023met,Calmet:2024tgm,Calmet:2024tyu}. We explicitly construct the quantum state of the gravity field, which was already discussed in general terms in  \cite{Calmet:2021stu}.

In QED, Barnich showed that the Coulomb field can be understood as a coherent state of temporal and longitudinal photons. The crucial mechanism is that the external charge shifts Gauss's law, and therefore shifts the BRST (see e.g. \cite{Henneaux:1994lbw} for a review) charge itself. The new BRST vacuum is annihilated by a shifted null annihilation operator, and that vacuum is a coherent state built from degrees of freedom in the unphysical sector of non-propagating U(1) modes.

The linearized gravity problem is closely analogous but not identical. The role of Gauss's law is now played by the linearized Hamiltonian and momentum constraints. For a static mass source, only the scalar Hamiltonian constraint is shifted. The static Newton field therefore lives in the \emph{non-radiative scalar constraint sector}, not in the propagating transverse-traceless graviton sector. That is why the naive idea of shifting free on-shell graviton operators fails: free radiative modes are the wrong degrees of freedom for a time-independent $1/r$ field.

The goal of the paper is to derive the gravity state in the following sequence:
\begin{align*}
\text{sourced constraints} &\Longrightarrow \text{sourced BRST charge} \\&\Longrightarrow \text{shifted null annihilator in the scalar sector} \\&\Longrightarrow \text{coherent BRST state describing the Newton field}.
\end{align*}

Throughout, we work in linearized gravity around Minkowski spacetime, with an external conserved static source. We regulate the point mass at intermediate stages by a smooth profile $f_a(\mathbf x)$ satisfying
\begin{equation}
\int d^3x\, f_a(\mathbf x)=1,
\qquad
\rho(\mathbf x)=M f_a(\mathbf x),
\end{equation}
so that the point-source limit is recovered as $f_a(\mathbf x)\to \delta^{(3)}(\mathbf x)$.

\section{Barnich's QED construction}

Here we review the BRST construction in electrodynamics.

\subsection{External charge and sourced Gauss law}

Consider Maxwell theory coupled to a static point charge $Q$ located at the origin. The action is
\begin{equation}
S_Q=\int d^4x\,\left[-\frac14 F_{\mu\nu}F^{\mu\nu}-j^\mu A_\mu\right],
\qquad
j^\mu=\delta^\mu_0\,Q\,\delta^{(3)}(\mathbf x).
\end{equation}

In the Hamiltonian formulation, the secondary constraint is Gauss's law. In the presence of the source it becomes
\begin{equation}
\phi^Q_2(\mathbf x)=-\partial_i \pi^i(\mathbf x)+j^0(\mathbf x)=0.
\end{equation}
This is the first key structural point: the source does not first show up as a classical field configuration; it shows up \emph{inside the constraint}. Barnich's BRST charge therefore becomes a sourced BRST charge.

\subsection{Barnich's sourced BRST charge}

Barnich writes the sourced BRST operator in momentum space as
\begin{equation}
\Omega_Q
=
\int d^3k\,
\Big[c^\dagger(\mathbf k)\big(a(\mathbf k)-q(\mathbf k)\big)
+\big(a^\dagger(\mathbf k)-q(\mathbf k)\big)c(\mathbf k)\Big],
\label{eq:BarnichBRST}
\end{equation}
with
\begin{equation}
q(\mathbf k)=\frac{Q}{(2\pi)^{3/2}\sqrt{2}\,k^{3/2}}.
\end{equation}
Here $a(\mathbf k)$ is a null annihilation operator built from the longitudinal and temporal oscillators, and $c(\mathbf k)$ is the ghost operator. The ordinary free BRST vacuum is no longer annihilated by \eqref{eq:BarnichBRST}, because the source has shifted the operator $a$ to $a-q$.

Barnich therefore defines the shifted annihilator
\begin{equation}
a_Q(\mathbf k):=a(\mathbf k)-q(\mathbf k),
\end{equation}
and the new vacuum by the condition
\begin{equation}
a_Q(\mathbf k)\,\ket{0}_Q=0.
\end{equation}
The new vacuum is a coherent state built from the null creation operator $b^\dagger(\mathbf k)$ conjugate to $a(\mathbf k)$.

\subsection{The null pair and the coherent-state solution}

The unphysical sector contains one positive-norm and one negative-norm oscillator. Barnich packages them into a null pair $a,b$ satisfying
\begin{equation}
[a(\mathbf k),b^\dagger(\mathbf k')]=\delta^{(3)}(\mathbf k-\mathbf k'),
\qquad
[a,a^\dagger]=[b,b^\dagger]=0.
\end{equation}

Now consider the candidate coherent state
\begin{equation}
\ket{0}_Q
=
\exp\!\left[\int d^3k\, q(\mathbf k) b^\dagger(\mathbf k)\right]\ket{0}.
\label{eq:QEDcoherentstate}
\end{equation}
We check explicitly that it solves the shifted annihilation condition. Let
\begin{equation}
X:=\int d^3k\, q(\mathbf k)b^\dagger(\mathbf k).
\end{equation}
Then, using the Baker--Campbell--Hausdorff identity,
\begin{equation}
a(\mathbf p)e^X=e^X\left(a(\mathbf p)+[a(\mathbf p),X]+\frac12 [[a(\mathbf p),X],X]+\cdots\right).
\end{equation}
Because $[a(\mathbf p),b^\dagger(\mathbf k)]=\delta^{(3)}(\mathbf p-\mathbf k)$, we have
\begin{equation}
[a(\mathbf p),X]=q(\mathbf p),
\end{equation}
and the next commutator vanishes because $q(\mathbf p)$ is a c-number. Therefore
\begin{equation}
a(\mathbf p)e^X=e^X(a(\mathbf p)+q(\mathbf p)).
\end{equation}
Multiplying both sides by $\ket{0}$ and using $a(\mathbf p)\ket{0}=0$, we get
\begin{equation}
(a(\mathbf p)-q(\mathbf p))\ket{0}_Q=0.
\end{equation}
So \eqref{eq:QEDcoherentstate} is indeed the shifted BRST vacuum.

\subsection{Why this matters for gravity}

The entire gravity construction below is an adaptation of this logic. We will need four ingredients:
\begin{enumerate}[leftmargin=1.6em]
  \item a sourced first-class constraint,
  \item a BRST charge in which that sourced constraint appears,
  \item a null annihilation operator $A(\mathbf k)$ adapted to the relevant sector of the constraint,
  \item a conjugate null creator $B^\dagger(\mathbf k)$ that generates the coherent shift.
\end{enumerate}
The only nontrivial new feature in gravity is that the relevant constraint sector is the \emph{scalar Hamiltonian constraint}, not a radiative spin-2 mode.

\section{Linearized gravity with source, gauge fixing, and BRST symmetry}

\subsection{Field variables and trace reversal}

We write
\begin{equation}
g_{\mu\nu}=\eta_{\mu\nu}+h_{\mu\nu},
\qquad
\eta_{\mu\nu}=\mathrm{diag}(-1,+1,+1,+1),
\qquad
|h_{\mu\nu}|\ll 1.
\end{equation}
Define the trace and trace-reversed field by
\begin{equation}
h=\eta^{\mu\nu}h_{\mu\nu},
\qquad
\bar h_{\mu\nu}=h_{\mu\nu}-\frac12\eta_{\mu\nu}h.
\end{equation}
Then
\begin{equation}
\bar h=\eta^{\mu\nu}\bar h_{\mu\nu}=-h,
\qquad
h_{\mu\nu}=\bar h_{\mu\nu}-\frac12\eta_{\mu\nu}\bar h.
\end{equation}

\subsection{Source coupling}

The linear coupling to an external conserved source is
\begin{equation}
S_{\text{int}}=\frac12\int d^4x\, h_{\mu\nu}T^{\mu\nu},
\qquad
\partial_\mu T^{\mu\nu}=0.
\end{equation}
For the regulated static mass source used in this paper we take
\begin{equation}
T^{00}(t,\mathbf x)=\rho(\mathbf x)=M f_a(\mathbf x),
\qquad
T^{0i}=0,
\qquad
T^{ij}=0.
\end{equation}

\subsection{Gauge fixing and ghosts}

The de Donder gauge condition is
\begin{equation}
C_\nu:=\partial^\mu \bar h_{\mu\nu}=0.
\end{equation}
Introduce a Nakanishi--Lautrup auxiliary field $B_\mu$ and ghosts $c_\mu,\bar c_\mu$. A convenient gauge-fixed action is
\begin{equation}
S=S_{\text{FP}}^{(2)}[h]
+\int d^4x\left[
B^\nu C_\nu+\frac\alpha2 B^\nu B_\nu
+i\bar c^\nu \Box c_\nu
+\frac12 h_{\mu\nu}T^{\mu\nu}
\right].
\end{equation}
The linearized BRST transformations are
\begin{equation}
s h_{\mu\nu}=\partial_\mu c_\nu+\partial_\nu c_\mu,
\qquad
s c_\mu=0,
\qquad
s\bar c_\mu=iB_\mu,
\qquad
s B_\mu=0.
\end{equation}
where $s$ is the BRST differential. With these conventions the sign of the ghost term is fixed: since $sC_\nu=\Box c_\nu$, the variation of $B^\nu C_\nu$ is canceled by the variation of $+i\bar c^\nu\Box c_\nu$.

\subsection{Explicit proof of BRST invariance of the source term}

This proof is very short, but it is worth writing out explicitly because it explains why the source shifts the graviton sector while leaving the ghost vacuum untouched.

Acting on the source term,
\begin{align}
s S_{\text{int}}
&=\frac12\int d^4x\, (\partial_\mu c_\nu+\partial_\nu c_\mu)T^{\mu\nu} \\
&=\int d^4x\, (\partial_\mu c_\nu)T^{\mu\nu}
\qquad\text{(because $T^{\mu\nu}=T^{\nu\mu}$)} \\
&=-\int d^4x\, c_\nu\,\partial_\mu T^{\mu\nu}
\qquad\text{(integrate by parts and drop the boundary term)} \\
&=0
\qquad\text{(because the source is conserved).}
\end{align}
So the source term is BRST invariant. In particular, the source does \emph{not} act as a source for the ghost sector. That means the ghost part of the vacuum remains the same as in the source-free theory.

\subsection{Classical sourced equation in de Donder gauge}

In de Donder gauge the linearized Einstein equation becomes
\begin{equation}
\Box \bar h_{\mu\nu}=-16\pi G\,T_{\mu\nu},
\label{eq:deDsourceeq}
\end{equation}
with the sign convention appropriate to our metric signature and source normalization. For the static source only the $00$ component is nonzero. With the mostly-plus metric convention, $T_{00}=T^{00}=\rho$ and $\Box=\nabla^2$ on a static field, hence
\begin{equation}
\nabla^2 \bar h_{00}^{\cl}(\mathbf{x})=-16\pi G\,\rho( \mathbf{x}),
\qquad
\bar h_{0i}^{\cl}=0,
\qquad
\bar h_{ij}^{\cl}=0.
\end{equation}
Define the regulated Newton potential by
\begin{equation}
\Phi_N(\mathbf x):=-G M\int d^3y\,\frac{f_a(\mathbf y)}{|\mathbf x-\mathbf y|}.
\label{eq:regulatedPhi}
\end{equation}
Using
\begin{equation}
\nabla^2\left(-\frac{1}{|\mathbf x-\mathbf y|}\right)=4\pi\delta^{(3)}(\mathbf x-\mathbf y),
\end{equation}
we obtain
\begin{equation}
\nabla^2\Phi_N(\mathbf x)=4\pi G\rho(\mathbf x).
\label{eq:PoissonPhi}
\end{equation}
Comparing with the $00$ equation above,
\begin{equation}
\bar h_{00}^{\cl}(\mathbf x)=-4\Phi_N(\mathbf x).
\end{equation}
The corresponding $h_{\mu\nu}$ components are
\begin{equation}
h_{00}^{\cl}=-2\Phi_N,
\qquad
h_{ij}^{\cl}=-2\Phi_N\,\delta_{ij},
\qquad
h_{0i}^{\cl}=0.
\label{eq:Newtonmetriccomponents}
\end{equation}
All of this is classical. The question of this paper is: \emph{what is the corresponding quantum state, derived by BRST methods rather than by simply postulating the shift \eqref{eq:Newtonmetriccomponents}?}

\section{Canonical form and the sourced constraints}

\subsection{ADM split at linear order}

To see where the source enters in the BRST charge, we need the canonical constraints. Write the spacetime perturbation in $3+1$ form:
\begin{equation}
\gamma_{ij}:=h_{ij},
\qquad
n_i:=h_{0i},
\qquad
n:=\frac12 h_{00}.
\end{equation}
The variable $\gamma_{ij}$ carries the spatial metric perturbation, while $n$ and $n_i$ play the roles of linearized lapse and shift.

The canonical action has the schematic form
\begin{equation}
S=\int dt\,d^3x\,
\left[
\pi^{ij}\dot\gamma_{ij}
-p_n\dot n
-p^i\dot n_i
-H_{\text{FP}}^{(2)}
-n\,\mathcal H_M
-n_i\,\mathcal H^i_M
\right]+S_{\gh},
\label{eq:canonicalaction}
\end{equation}
where $\pi^{ij}$ is the momentum conjugate to $\gamma_{ij}$, the momenta $p_n,p^i$ enforce the primary constraints, and $\mathcal H_M$, $\mathcal H^i_M$ are the sourced secondary constraints.

\subsection{The sourced Hamiltonian and momentum constraints}

In the convention used in this paper, the unsourced scalar constraint is the linearized spatial Ricci scalar,
\begin{equation}
\mathcal H=\partial_i\partial_j\gamma_{ij}-\nabla^2\gamma,
\qquad
\gamma:=\delta^{ij}\gamma_{ij}.
\label{eq:Hconstraint}
\end{equation}
\begin{equation}
\mathcal H_i=-2\partial_j\pi_i{}^j.
\label{eq:Momconstraint}
\end{equation}
The external mass density shifts only the scalar Hamiltonian constraint:
\begin{equation}
\mathcal H_M=\mathcal H-16\pi G\rho,
\qquad
\mathcal H_i^M=\mathcal H_i.
\label{eq:sourcedconstraints}
\end{equation}

At this point the analogy with QED should be very clear. The charge density $j^0$ shifted Gauss's law. Here the mass density $\rho$ shifts the Hamiltonian constraint.

\subsection{Check that the Hamiltonian constraint reproduces the Newton equation}

For the static isotropic Newton configuration compatible with
\eqref{eq:Newtonmetriccomponents},
\begin{equation}
\gamma_{ij}=-2\Phi\,\delta_{ij}.
\end{equation}
Then
\begin{equation}
\gamma=\delta^{ij}\gamma_{ij}=-6\Phi,
\end{equation}
and
\begin{equation}
\partial_i\partial_j\gamma_{ij}
=
\partial_i\partial_j(-2\Phi\,\delta_{ij})
=
-2\nabla^2\Phi.
\end{equation}
Substituting into \eqref{eq:Hconstraint},
\begin{equation}
\mathcal H
=
-2\nabla^2\Phi-\nabla^2(-6\Phi)
=
4\nabla^2\Phi.
\label{eq:HconstraintPhi}
\end{equation}
Now impose the sourced constraint \eqref{eq:sourcedconstraints}:
\begin{equation}
0=\mathcal H_M=4\nabla^2\Phi-16\pi G\rho.
\end{equation}
Dividing by $4$ gives
\begin{equation}
\nabla^2\Phi=4\pi G\rho.
\end{equation}
This is exactly \eqref{eq:PoissonPhi}. So the Newton potential is encoded directly in the sourced Hamiltonian constraint.

This is the central structural observation. The Newton field is not being put in by hand; it is the solution to the sourced scalar constraint.

\section{The BFV-BRST charge for the sourced linearized theory}

\subsection{Minimal and nonminimal variables}

The lapse and shift variables $n,n_i$ have primary constraints $p_n\approx0$ and $p^i\approx0$. Conservation of these primary constraints gives the secondary constraints $\mathcal H_M\approx0$ and $\mathcal H_i^M\approx0$. In the linearized theory the first-class constraint algebra is abelian.

Introduce minimal ghosts $c,c_i$ for the secondary constraints, with conjugate ghost momenta $\mathcal P,\mathcal P_i$, and introduce independent minimal ghosts $C,C_i$ for the primary constraints, with conjugate ghost momenta $\Pi,\Pi_i$. The ghost-number assignments are
\begin{equation}
\operatorname{gh}(c)=\operatorname{gh}(c_i)=\operatorname{gh}(C)=\operatorname{gh}(C_i)=1,
\qquad
\operatorname{gh}(\mathcal P)=\operatorname{gh}(\mathcal P_i)=\operatorname{gh}(\Pi)=\operatorname{gh}(\Pi_i)=-1.
\end{equation}
The nonminimal antighost/Nakanishi--Lautrup variables used for gauge fixing may be added in the usual way, but they are not needed to display the source-dependent part of the BRST charge.

\subsection{The sourced BFV charge}

The minimal BFV-BRST operator is
\begin{equation}
\Omega_M
=
\int d^3x\,
\Big[
 c\,\mathcal H_M
 +c_i\,\mathcal H_i^M
 +C\,p_n
 +C_i\,p^i
\Big].
\label{eq:BFVcharge}
\end{equation}
This has ghost number $+1$. Because the linearized constraint algebra is abelian and the source is fixed c-number data, the charge is nilpotent:
\begin{equation}
\Omega_M^2=0.
\end{equation}
Equivalently, the graded Poisson bracket or graded commutator of \eqref{eq:BFVcharge} with itself vanishes. The important point is that the primary-constraint terms are multiplied by their own ghosts $C,C_i$, not by the ghost momenta conjugate to $c,c_i$.

The physical state condition is therefore
\begin{equation}
\Omega_M\ket{\Psi_M}=0.
\label{eq:physstatecond}
\end{equation}
In a fully explicit canonical treatment one also chooses a gauge-fixing fermion $\Psi$ and constructs the gauge-fixed Hamiltonian $H_\Psi=H+\{\Psi,\Omega_M\}$. For the present purpose, however, the essential piece is already visible in \eqref{eq:BFVcharge}: the source enters by shifting the Hamiltonian constraint itself.

\subsection{Why this already differs from the naive free-mode shift}

Equation \eqref{eq:BFVcharge} tells us what must be shifted. The shifted object is not a free radiative graviton annihilation operator. It is the operator content of the \emph{sourced scalar constraint}. This distinction matters because the static Newton solution is time independent and obeys a Poisson equation, whereas free radiative modes obey wave equations and evolve with factors $e^{-ikt}$.

That is why the correct Barnich-style question in gravity is not
\begin{equation*}
\text{``How do I shift a free graviton mode?''}
\end{equation*}
but rather
\begin{equation*}
\begin{gathered}
\text{``Which annihilation operator appears in the positive-frequency}\\
\text{part of the scalar constraint?''}
\end{gathered}
\end{equation*}
Once that operator is identified, the rest of the Barnich logic goes through exactly.

\section{Isolating the scalar constraint sector}

\subsection{Scalar decomposition of the spatial metric perturbation}

In Fourier space, a general scalar part of $\gamma_{ij}$ can be written as
\begin{equation}
\gamma_{ij}^S(\mathbf k)
=
\alpha(\mathbf k)\,\delta_{ij}
+\beta(\mathbf k)\left(\hat k_i\hat k_j-\frac13\delta_{ij}\right),
\qquad
\hat k_i:=\frac{k_i}{k}.
\label{eq:scalarDecomp}
\end{equation}
The first term is isotropic. The second is the traceless scalar built from $\mathbf k$.

Let us compute the Fourier transform of the Hamiltonian constraint explicitly. In momentum space, every spatial derivative $\partial_i$ becomes multiplication by $ik_i$, and therefore
\begin{equation}
\partial_i\partial_j\gamma_{ij}(\mathbf x)
\longrightarrow
-k_i k_j\gamma_{ij}(\mathbf k),
\qquad
\nabla^2\gamma(\mathbf x)
\longrightarrow
-k^2\gamma(\mathbf k).
\end{equation}
From \eqref{eq:scalarDecomp}, the trace is
\begin{equation}
\gamma(\mathbf k)=\delta^{ij}\gamma_{ij}^S(\mathbf k)=3\alpha(\mathbf k),
\end{equation}
because the traceless term has zero trace. Also,
\begin{align}
k_i k_j\gamma_{ij}^S(\mathbf k)
&=k_i k_j\left[\alpha\,\delta_{ij}+\beta\left(\hat k_i\hat k_j-\frac13\delta_{ij}\right)\right] \\
&=\alpha\,k_i k_i+\beta\left[k_i k_j\hat k_i\hat k_j-\frac13 k_i k_j\delta_{ij}\right] \\
&=\alpha k^2+\beta\left[k^2-\frac13 k^2\right] \\
&=\alpha k^2+\frac23\beta k^2.
\end{align}
Substituting into \eqref{eq:Hconstraint},
\begin{align}
\mathcal H(\mathbf k)
&=\big(-k_i k_j\gamma_{ij}(\mathbf k)\big)-\big(-k^2\gamma(\mathbf k)\big) \\
&=-k_i k_j\gamma_{ij}(\mathbf k)+k^2\gamma(\mathbf k) \\
&=-\left(\alpha k^2+\frac23\beta k^2\right)+3\alpha k^2 \\
&=2\alpha k^2-\frac23\beta k^2.
\end{align}
It is convenient to package the specific scalar combination entering the constraint into a single variable,
\begin{equation}
\chi(\mathbf k):=-\alpha(\mathbf k)+\frac13\beta(\mathbf k).
\label{eq:chiDef}
\end{equation}
Then the constraint becomes
\begin{equation}
\mathcal H(\mathbf k)=-2k^2\chi(\mathbf k).
\label{eq:Hchi}
\end{equation}
This is the unique scalar combination that the Hamiltonian constraint sees.

\subsection{The isotropic Newton configuration as a special case}

For the isotropic Newton field, $\beta=0$ and $\alpha=-2\Phi$. Therefore
\begin{equation}
\chi=2\Phi,
\qquad
\mathcal H(\mathbf k)=-2k^2(2\Phi)= -4k^2\Phi(\mathbf k).
\label{eq:HkPhi}
\end{equation}
Since the sourced constraint says $\mathcal H(\mathbf k)=16\pi G\rho(\mathbf k)$, we recover
\begin{equation}
-4k^2\Phi(\mathbf k)=16\pi G\rho(\mathbf k),
\qquad\text{so}\qquad
\Phi(\mathbf k)=-\frac{4\pi G\rho(\mathbf k)}{k^2}.
\label{eq:PhiFourier}
\end{equation}
This is the momentum-space Newton potential.

\section{The null scalar oscillator basis}

\subsection{Canonical normalization of the covariant graviton oscillators}

The scalar null pair should not be introduced by fiat. It follows from the canonical oscillator algebra of the gauge-fixed linearized graviton. To make the dimensions explicit, write the dimensionless metric perturbation as
\begin{equation}
h_{\mu\nu}=\kappa\,\psi_{\mu\nu},
\qquad
\kappa^2=32\pi G,
\label{eq:kappaPsiDef}
\end{equation}
where $\psi_{\mu\nu}$ is the canonically normalized spin-2 field. In de Donder gauge, the positive-frequency part of the spatial field may be expanded as
\begin{equation}
\psi^{(+)}_{ij}(t,\mathbf x)
=
\int \frac{d^3k}{(2\pi)^3}\,\frac{1}{\sqrt{2k}}\,
 a_{ij}(\mathbf k)e^{-ikt+i\mathbf k\cdot\mathbf x}.
\label{eq:psiOscExpansion}
\end{equation}
The spatial oscillator algebra is
\begin{equation}
[a_{ij}(\mathbf k),a^\dagger_{\ell m}(\mathbf k')]
=(2\pi)^3\delta^{(3)}(\mathbf k-\mathbf k')\,
P_{ij,\ell m},
\label{eq:spatialOscAlg}
\end{equation}
with
\begin{equation}
P_{ij,\ell m}
=\frac{1}{2}\left(
\delta_{i\ell}\delta_{jm}
+\delta_{im}\delta_{j\ell}
-\delta_{ij}\delta_{\ell m}
\right).
\label{eq:spatialDewittMetric}
\end{equation}
Equivalently, for two symmetric spatial tensors $s_{ij}$ and $t_{ij}$,
\begin{equation}
[s^{ij}a_{ij}(\mathbf k),t^{\ell m}a^\dagger_{\ell m}(\mathbf k')]
=(2\pi)^3\delta^{(3)}(\mathbf k-\mathbf k')\,
\langle s,t\rangle_{\rm DW},
\label{eq:DewittCommutator}
\end{equation}
where
\begin{equation}
\langle s,t\rangle_{\rm DW}
=s^{ij}t_{ij}-\frac{1}{2}s^i{}_{i}t^j{}_{j}.
\label{eq:DewittInnerProduct}
\end{equation}
This indefinite DeWitt metric is the source of the scalar null pair. No additional assumption about positive- and negative-norm scalar oscillators is needed.

\subsection{Scalar projectors and the null direction selected by the constraint}

For fixed nonzero $\mathbf k$, define
\begin{equation}
\ell_{ij}:=\hat k_i\hat k_j,
\qquad
\theta_{ij}:=\delta_{ij}-\hat k_i\hat k_j.
\label{eq:thetaellDef}
\end{equation}
The traces and contractions are
\begin{equation}
\theta^i{}_i=2,
\qquad
\ell^i{}_i=1,
\qquad
\theta_{ij}\ell^{ij}=0,
\qquad
\theta_{ij}\theta^{ij}=2,
\qquad
\ell_{ij}\ell^{ij}=1.
\end{equation}
Using \eqref{eq:DewittInnerProduct}, one obtains
\begin{equation}
\langle\theta,\theta\rangle_{\rm DW}=0,
\qquad
\langle\ell,\ell\rangle_{\rm DW}=\frac{1}{2},
\qquad
\langle\theta,\ell\rangle_{\rm DW}=-1.
\label{eq:thetaellDW}
\end{equation}
Thus the transverse scalar tensor $\theta_{ij}$ is itself a null polarization of the gauge-fixed spatial graviton sector.

Now compute the positive-frequency part of the linearized Hamiltonian constraint. With the convention used in \eqref{eq:Hconstraint},
\begin{equation}
\mathcal H=\partial_i\partial_j h_{ij}-\nabla^2 h^i{}_i.
\label{eq:HminusRagain}
\end{equation}
Since $h_{ij}=\kappa\psi_{ij}$, equations \eqref{eq:psiOscExpansion} and \eqref{eq:HminusRagain} give
\begin{align}
\mathcal H^{(+)}(\mathbf k)
&=\frac{\kappa}{\sqrt{2k}}\left(-k_i k_j+k^2\delta_{ij}\right)a_{ij}(\mathbf k) \\
&=\frac{\kappa k^{3/2}}{\sqrt2}\,\theta^{ij}a_{ij}(\mathbf k).
\label{eq:HplusTheta}
\end{align}
It is therefore natural to define the active scalar annihilation operator by
\begin{equation}
A(\mathbf k):=\theta^{ij}a_{ij}(\mathbf k).
\label{eq:AfromTheta}
\end{equation}
Then
\begin{equation}
\mathcal H^{(+)}(\mathbf k)
=\frac{\kappa k^{3/2}}{\sqrt2}\,A(\mathbf k).
\label{eq:HplusAderived}
\end{equation}
This replaces the earlier normalization-by-definition. The factor of $\kappa$ is required by canonical normalization and is the origin of the physically correct $\sqrt G$ scaling of the coherent-state amplitude.

Because $\theta$ is null in the DeWitt metric,
\begin{equation}
[A(\mathbf k),A^\dagger(\mathbf k')]=0.
\label{eq:AnullDerived}
\end{equation}
Thus the operator that appears in the positive-frequency part of the scalar Hamiltonian constraint is a null annihilation operator.

\subsection{The conjugate null partner}

We now construct a conjugate null oscillator $B(\mathbf k)$ satisfying
\begin{equation}
[A(\mathbf k),B^\dagger(\mathbf k')]
=(2\pi)^3\delta^{(3)}(\mathbf k-\mathbf k'),
\qquad
[B(\mathbf k),B^\dagger(\mathbf k')]=0.
\label{eq:ABdesiredDerived}
\end{equation}
Let
\begin{equation}
B(\mathbf k)
=-\left(\frac{1}{4}\theta^{ij}+\ell^{ij}\right)a_{ij}(\mathbf k).
\label{eq:BfromThetaell}
\end{equation}
Then, using \eqref{eq:AfromTheta}, \eqref{eq:DewittCommutator}, and \eqref{eq:thetaellDW},
\begin{align}
[A(\mathbf k),B^\dagger(\mathbf k')]
&=-(2\pi)^3\delta^{(3)}(\mathbf k-\mathbf k')
\left\langle \theta,\frac{1}{4}\theta+\ell\right\rangle_{\rm DW} \\
&=-(2\pi)^3\delta^{(3)}(\mathbf k-\mathbf k')
\left(0-1\right) \\
&=(2\pi)^3\delta^{(3)}(\mathbf k-\mathbf k').
\end{align}
Similarly,
\begin{align}
\left\langle \frac{1}{4}\theta+\ell,\frac{1}{4}\theta+\ell\right\rangle_{\rm DW}
&=\frac{1}{16}\langle\theta,\theta\rangle_{\rm DW}
+\frac{1}{2}\langle\theta,\ell\rangle_{\rm DW}
+\langle\ell,\ell\rangle_{\rm DW} \\
&=0-\frac{1}{2}+\frac{1}{2}=0,
\end{align}
so
\begin{equation}
[B(\mathbf k),B^\dagger(\mathbf k')]=0.
\end{equation}
Hence the scalar constraint sector contains the null pair
\begin{equation}
[A,A^\dagger]=0,
\qquad
[B,B^\dagger]=0,
\qquad
[A(\mathbf k),B^\dagger(\mathbf k')]
=(2\pi)^3\delta^{(3)}(\mathbf k-\mathbf k').
\label{eq:ABalgebraDerived}
\end{equation}
This is the gravity analogue of Barnich's longitudinal-temporal null pair in QED, now obtained directly from the gauge-fixed canonical graviton algebra.

\subsection{Relation to positive- and negative-norm scalar oscillators}

The previous algebra can be put into the more familiar Barnich form by defining
\begin{equation}
a_+(\mathbf k):=\frac{A(\mathbf k)+B(\mathbf k)}{\sqrt2},
\qquad
a_-(\mathbf k):=\frac{A(\mathbf k)-B(\mathbf k)}{\sqrt2}.
\label{eq:aplusminusFromAB}
\end{equation}
Then \eqref{eq:ABalgebraDerived} implies
\begin{equation}
[a_+(\mathbf k),a_+^\dagger(\mathbf k')]
=(2\pi)^3\delta^{(3)}(\mathbf k-\mathbf k'),
\qquad
[a_-(\mathbf k),a_-^\dagger(\mathbf k')]
=-(2\pi)^3\delta^{(3)}(\mathbf k-\mathbf k'),
\end{equation}
with mixed commutators vanishing. Conversely,
\begin{equation}
A=\frac{a_++a_-}{\sqrt2},
\qquad
B=\frac{a_+-a_-}{\sqrt2}.
\end{equation}
Thus the positive- and negative-norm scalar oscillators are not an independent postulate. They are just the diagonal basis of the two-dimensional scalar subspace spanned by $\theta_{ij}$ and $\ell_{ij}$ under the DeWitt metric.

\subsection{Hermitian constraint operator}

With the conventions above, the full Hermitian scalar constraint operator is
\begin{equation}
\mathcal H(\mathbf k)
=\frac{\kappa k^{3/2}}{\sqrt2}\,
\Big(A(\mathbf k)+A^\dagger(-\mathbf k)\Big).
\label{eq:HfullA}
\end{equation}
For a static source the coherent amplitudes may be chosen real and even in $\mathbf k$, so the distinction between $\mathbf k$ and $-\mathbf k$ will not affect the following formulae.

\section{Solving the sourced BRST condition}

\subsection{The scalar part of the sourced BRST charge}

Once the scalar null pair has been derived, the scalar part of the sourced BRST charge takes the Barnich form
\begin{equation}
\Omega_M^{(S)}
=
\int \frac{d^3k}{(2\pi)^3}\,
\Big[
 c^\dagger(\mathbf k)\big(A(\mathbf k)-q_M(\mathbf k)\big)
 +\big(A^\dagger(\mathbf k)-q_M(\mathbf k)\big)c(\mathbf k)
\Big],
\label{eq:gravityscalarBRST}
\end{equation}
where the c-number $q_M(\mathbf k)$ is fixed by the sourced constraint. The full BRST state will be obtained by tensoring this scalar-sector state with the vacuum of the vector, transverse-traceless, ghost, and nonminimal sectors.

\subsection{Determining \texorpdfstring{$q_M(\mathbf k)$}{qM(k)} from the sourced constraint}

By construction, the coherent BRST state should satisfy
\begin{equation}
\expval{\mathcal H(\mathbf k)}_{M,a}=16\pi G\rho(\mathbf k)
=16\pi G M\,\tilde f_a(\mathbf k),
\label{eq:expectationconstraint}
\end{equation}
where the subscript reminds us that the source is regulated by $f_a$.

Using \eqref{eq:HfullA}, the left-hand side is
\begin{equation}
\expval{\mathcal H(\mathbf k)}_{M,a}
=
\frac{\kappa k^{3/2}}{\sqrt2}\,
\expval{A(\mathbf k)+A^\dagger(-\mathbf k)}_{M,a}.
\end{equation}
In the coherent state constructed below, the expectation values of $A$ and $A^\dagger$ are both equal to $q_M$ for real even amplitudes. Therefore
\begin{equation}
\expval{\mathcal H(\mathbf k)}_{M,a}
=\sqrt2\,\kappa k^{3/2}q_M(\mathbf k).
\end{equation}
Comparing with \eqref{eq:expectationconstraint}, one obtains
\begin{equation}
q_M(\mathbf k)
=\frac{16\pi G M\,\tilde f_a(\mathbf k)}{\sqrt2\,\kappa k^{3/2}}
=2\sqrt{\pi G}\,\frac{M\tilde f_a(\mathbf k)}{k^{3/2}}.
\label{eq:qMresult}
\end{equation}
The exact sign of $q_M$ follows the sign convention chosen for the constraint and for $A$. The important invariant point is the scaling $q_M\sim \sqrt G M/k^{3/2}$, not $GM/k^{3/2}$.

\subsection{The coherent-state solution}

Define the shifted annihilator
\begin{equation}
A_M(\mathbf k):=A(\mathbf k)-q_M(\mathbf k).
\end{equation}
We now solve the condition
\begin{equation}
A_M(\mathbf k)\ket{M;a}_S=0
\qquad\text{for all $\mathbf k$.}
\label{eq:shiftedannihilation}
\end{equation}
Take the scalar vacuum $\ket{0}_S$ annihilated by $A(\mathbf k)$ and $B(\mathbf k)$, and define
\begin{equation}
\ket{M;a}_S
:=
\exp\!\left[\int \frac{d^3k}{(2\pi)^3}\,q_M(\mathbf k)B^\dagger(\mathbf k)\right]\ket{0}_S.
\label{eq:gravitycoherentstate}
\end{equation}
Exactly as in the QED warm-up,
\begin{equation}
A(\mathbf p)e^X=e^X(A(\mathbf p)+q_M(\mathbf p)),
\qquad
X:=\int \frac{d^3k}{(2\pi)^3}\,q_M(\mathbf k)B^\dagger(\mathbf k),
\end{equation}
because \eqref{eq:ABalgebraDerived} gives $[A(\mathbf p),X]=q_M(\mathbf p)$ and the second commutator vanishes. Therefore
\begin{equation}
(A(\mathbf p)-q_M(\mathbf p))\ket{M;a}_S=0.
\end{equation}
So \eqref{eq:gravitycoherentstate} is the desired scalar-sector BRST coherent state.

The full sourced BRST state is then
\begin{equation}
\ket{M;a}_{\BRST}
=
\ket{M;a}_S
\otimes \ket{0}_V
\otimes \ket{0}_{\TT}
\otimes \ket{0}_{\gh}
\otimes \ket{0}_{\text{nonmin}}.
\label{eq:fullstate}
\end{equation}
This makes the physical interpretation precise: the source shifts the scalar constraint sector while the propagating transverse-traceless gravitons remain in their vacuum.

\section{Computing the Newton potential from the BRST state}

\subsection{Expectation value of the constraint}

By construction,
\begin{equation}
\expval{\mathcal H(\mathbf k)}_{M,a}=16\pi G M\,\tilde f_a(\mathbf k).
\end{equation}
The Fourier-space relation \eqref{eq:HkPhi} for the isotropic Newton sector gives
\begin{equation}
\mathcal H(\mathbf k)=-4k^2\Phi(\mathbf k).
\end{equation}
Therefore
\begin{equation}
-4k^2\expval{\Phi(\mathbf k)}_{M,a}=16\pi G M\,\tilde f_a(\mathbf k),
\end{equation}
and hence
\begin{equation}
\expval{\Phi(\mathbf k)}_{M,a}
=-\frac{4\pi G M\,\tilde f_a(\mathbf k)}{k^2}.
\label{eq:Phiexpectk}
\end{equation}
This is exactly the regulated momentum-space Newton potential.

\subsection{Transforming back to position space}

Using the Fourier convention
\begin{equation}
f(\mathbf x)=\int \frac{d^3k}{(2\pi)^3}\,e^{i\mathbf k\cdot \mathbf x}\,\tilde f(\mathbf k),
\qquad
\tilde f(\mathbf k)=\int d^3x\,e^{-i\mathbf k\cdot\mathbf x}f(\mathbf x),
\label{eq:Fourierconv}
\end{equation}
we have the standard identity
\begin{equation}
\int \frac{d^3k}{(2\pi)^3}\,\frac{e^{i\mathbf k\cdot(\mathbf x-\mathbf y)}}{k^2}=\frac{1}{4\pi |\mathbf x-\mathbf y|}.
\label{eq:k2identity}
\end{equation}
Applying this to \eqref{eq:Phiexpectk},
\begin{align}
\expval{\Phi(\mathbf x)}_{M,a}
&=\int \frac{d^3k}{(2\pi)^3}\,e^{i\mathbf k\cdot\mathbf x}\expval{\Phi(\mathbf k)}_{M,a} \\
&=-4\pi G M\int \frac{d^3k}{(2\pi)^3}\,e^{i\mathbf k\cdot\mathbf x}\,\frac{\tilde f_a(\mathbf k)}{k^2} \\
&=-4\pi G M\int d^3y\,f_a(\mathbf y)\int \frac{d^3k}{(2\pi)^3}\,\frac{e^{i\mathbf k\cdot(\mathbf x-\mathbf y)}}{k^2} \\
&=-G M\int d^3y\,\frac{f_a(\mathbf y)}{|\mathbf x-\mathbf y|}.
\end{align}
This is exactly \eqref{eq:regulatedPhi}.

In the point-source limit $f_a\to\delta^{(3)}$, we obtain
\begin{equation}
\expval{\Phi(\mathbf x)}_M\to -\frac{GM}{r},
\qquad
r:=|\mathbf x|.
\end{equation}
So the BRST coherent state \eqref{eq:fullstate} carries the Newton field.

\subsection{Metric expectation values}

Using \eqref{eq:Newtonmetriccomponents}, we therefore have
\begin{equation}
\expval{h_{00}(\mathbf x)}_{M,a}=-2\expval{\Phi(\mathbf x)}_{M,a},
\qquad
\expval{h_{ij}(\mathbf x)}_{M,a}=-2\expval{\Phi(\mathbf x)}_{M,a}\,\delta_{ij},
\qquad
\expval{h_{0i}(\mathbf x)}_{M,a}=0.
\end{equation}
Meanwhile,
\begin{equation}
\expval{h_{ij}^{\TT}(\mathbf x)}_{M,a}=0,
\end{equation}
because the TT sector was not shifted at all. This is an operator-theoretic way of stating that Newton's law comes from the non-radiative scalar constraint sector, not from a cloud of TT gravitons.

\section{Why this is not ``just shifting the classical background''}

At first sight the final answer may look similar to the familiar displaced-vacuum picture of a quadratic field theory. Indeed, once the classical sourced solution $\bar h_{\mu\nu}^{\cl}$ is known, one may write a formal translation operator
\begin{equation}
U_M
=
\exp\!\left[-i\int d^3x\,\bar h_{\mu\nu}^{\cl}(\mathbf x;M)\,\Pi^{\mu\nu}_{\bar h}(\mathbf x)\right],
\end{equation}
and then the source-dependent vacuum can be written as
\begin{equation}
\ket{\Omega_M}=U_M\ket{0}.
\label{eq:shiftedvacuum}
\end{equation}

However, that is not how we derived the state in this paper. Our logic was:
\begin{enumerate}[leftmargin=1.6em]
  \item identify the sourced Hamiltonian constraint,
  \item build the sourced BRST charge,
  \item solve the sourced BRST annihilation condition,
  \item obtain the coherent state in the scalar null sector,
  \item and only then recognize that the result is equivalent to \eqref{eq:shiftedvacuum}.
\end{enumerate}

This order matters conceptually. The displaced-vacuum formula \eqref{eq:shiftedvacuum} is a useful repackaging of the answer, but it does not explain \emph{which sector} has been shifted or \emph{why} the Newton field is not made of radiative gravitons. The BRST derivation does explain that, because it reveals that the source sits directly in the scalar Hamiltonian constraint.

\section{Point-source limit and generalized states}

For smooth regulator $f_a$, the state \eqref{eq:fullstate} is a perfectly good Gaussian-coherent state of the linearized theory. In the ideal point-source limit, however, two familiar singular features appear:
\begin{enumerate}[leftmargin=1.6em]
  \item a short-distance singularity at $r=0$, coming from the behavior $\Phi\sim -GM/r$;
  \item a long-range infrared tail, because the field falls only as $1/r$.
\end{enumerate}
These features mean that the exact point-mass state is best regarded as a generalized coherent state or algebraic state rather than as a normalizable vector in the original free Fock space. This is the gravity analogue of the infrared subtleties already familiar from QED.

\section{Summary of Results}

Let us collect the central formulae in one place.

\subsection*{Step 1: sourced constraint}
For a static regulated mass density $\rho(\mathbf x)=M f_a(\mathbf x)$,
\begin{equation}
\mathcal H_M(\mathbf x)=\mathcal H(\mathbf x)-16\pi G\rho(\mathbf x),
\qquad
\mathcal H(\mathbf x)=\partial_i\partial_j\gamma_{ij}-\nabla^2\gamma.
\end{equation}

\subsection*{Step 2: sourced BRST charge}
The corresponding BFV charge is
\begin{equation}
\Omega_M
=
\int d^3x\,
\Big[
 c\,\mathcal H_M
 +c_i\,\mathcal H_i^M
 +C\,p_n
 +C_i\,p^i
\Big].
\end{equation}

\subsection*{Step 3: scalar null mode}
The scalar null pair is derived from the gauge-fixed canonical graviton algebra. With $h_{ij}=\kappa\psi_{ij}$ and $\kappa^2=32\pi G$,
\begin{equation}
A(\mathbf k)=\theta^{ij}a_{ij}(\mathbf k),
\qquad
B(\mathbf k)=-\left(\frac{1}{4}\theta^{ij}+\ell^{ij}\right)a_{ij}(\mathbf k),
\end{equation}
where $\theta_{ij}=\delta_{ij}-\hat k_i\hat k_j$ and $\ell_{ij}=\hat k_i\hat k_j$. These satisfy
\begin{equation}
[A(\mathbf k),B^\dagger(\mathbf k')]
=(2\pi)^3\delta^{(3)}(\mathbf k-\mathbf k'),
\qquad
[A,A^\dagger]=[B,B^\dagger]=0,
\end{equation}
and the constraint operator is
\begin{equation}
\mathcal H^{(+)}(\mathbf k)
=\frac{\kappa k^{3/2}}{\sqrt2}\,A(\mathbf k).
\end{equation}

\subsection*{Step 4: coherent amplitude}
The sourced BRST condition fixes
\begin{equation}
q_M(\mathbf k)=2\sqrt{\pi G}\,\frac{M\tilde f_a(\mathbf k)}{k^{3/2}},
\end{equation}
up to the overall sign convention.

\subsection*{Step 5: Newton state}
The scalar-sector state is
\begin{equation}
\ket{M;a}_S
=
\exp\!\left[\int \frac{d^3k}{(2\pi)^3}\,q_M(\mathbf k)B^\dagger(\mathbf k)\right]\ket{0}_S,
\end{equation}
and the full source-dependent BRST state is
\begin{equation}
\ket{M;a}_{\BRST}
=
\ket{M;a}_S
\otimes \ket{0}_V
\otimes \ket{0}_{\TT}
\otimes \ket{0}_{\gh}
\otimes \ket{0}_{\text{nonmin}}
.
\end{equation}

\subsection*{Step 6: Newton potential}
This state satisfies
\begin{equation}
\expval{\Phi(\mathbf k)}_{M,a}
=-\frac{4\pi G M\,\tilde f_a(\mathbf k)}{k^2}
\end{equation}
and therefore
\begin{equation}
\expval{\Phi(\mathbf x)}_{M,a}
=-G M\int d^3y\,\frac{f_a(\mathbf y)}{|\mathbf x-\mathbf y|}
\to -\frac{GM}{r}
\quad (f_a\to\delta^{(3)}).
\end{equation}

This is the Barnich-style BRST derivation of the Newton state in linearized gravity.


\section{Dressing norm and corpuscular scaling}

The coherent state constructed above describes the static Newton field as a
displacement in the non-radiative scalar constraint sector of the unreduced BRST
Hilbert space. It is therefore important not to confuse this state with a
coherent state of propagating transverse-traceless gravitons. The
transverse-traceless sector remains in its vacuum,
\begin{equation}
\langle h^{\rm TT}_{ij}(x)\rangle_{M,a}=0 ,
\end{equation}
and the Newtonian \(1/r\) field is carried instead by the scalar
constraint-sector dressing.

Nevertheless, one may ask whether the coherent dressing has a useful
quantitative size. In ordinary positive-norm coherent states, the squared
coherent amplitude is naturally interpreted as a particle occupation number. In
the present case the relevant operators are null operators in an
indefinite-metric space, so no gauge-invariant graviton-number interpretation is
available. However, the integral of the squared displacement amplitude still
provides a useful diagnostic of the size of the unreduced dressing. We shall
call this quantity the dressing norm, or more cautiously the dressing
mode-count,
\begin{equation}
{\cal N}_{\rm dress}
:=
\int \frac{d^3 k}{(2\pi)^3}\, |q_M(k)|^2 .
\end{equation}
This is not an observable by itself. It depends on the choice of BRST
representative, the infrared regulator, and the ultraviolet smearing of the
source. Its role is only to measure the size of the coherent displacement needed
to reproduce the classical Newton field.

Using the amplitude derived above,
\begin{equation}
q_M(k)
=
\frac{2\sqrt{\pi G}\,M\,\widetilde f_a(k)}{k^{3/2}},
\end{equation}
we find
\begin{equation}
|q_M(k)|^2
=
\frac{4\pi G M^2 |\widetilde f_a(k)|^2}{k^3}.
\end{equation}
Therefore
\begin{equation}
{\cal N}_{\rm dress}
=
4\pi G M^2
\int \frac{d^3 k}{(2\pi)^3}
\frac{|\widetilde f_a(k)|^2}{k^3}.
\end{equation}
For a compact source whose profile suppresses momenta above
\(\Lambda \sim a^{-1}\), and for an infrared cutoff \(\mu \sim L^{-1}\), the
leading behaviour is
\begin{equation}
{\cal N}_{\rm dress}
\simeq
4\pi G M^2
\frac{4\pi}{(2\pi)^3}
\int_\mu^\Lambda \frac{dk}{k}.
\end{equation}
Thus
\begin{equation}
{\cal N}_{\rm dress}
\simeq
\frac{2G M^2}{\pi}
\log\left(\frac{\Lambda}{\mu}\right),
\end{equation}
up to convention-dependent numerical factors. The robust statement is therefore
the scaling
\begin{equation}
{\cal N}_{\rm dress}
\sim
G M^2
\log\left(\frac{\Lambda}{\mu}\right)
=
\left(\frac{M}{M_{\rm P}}\right)^2
\log\left(\frac{\Lambda}{\mu}\right),
\end{equation}
where \(M_{\rm P}^{-2}=G\) in the convention used here.

This logarithmic behaviour has a simple origin. The Newton potential behaves as
\(1/r\), so in momentum space the field scales as \(1/k^2\). The corresponding
coherent displacement scales as
\begin{equation}
q_M(k)
\sim
\frac{\sqrt{G}\,M}{k^{3/2}},
\end{equation}
and the measure \(d^3k\) then produces a logarithmic integral,
\begin{equation}
d^3 k\, |q_M(k)|^2
\sim
k^2 dk\, \frac{G M^2}{k^3}
\sim
G M^2 \frac{dk}{k}.
\end{equation}
The logarithmic infrared sensitivity is therefore a direct consequence of the
long-range nature of the Newton field.

For a compact object of size \(R\), a natural infrared cutoff is of order
\(\mu \sim R_{\rm IR}^{-1}\), where \(R_{\rm IR}\) is the scale at which the
asymptotic field is effectively measured or enclosed. The ultraviolet cutoff is
set either by the inverse size of the source, \(\Lambda \sim a^{-1}\), or by the
scale at which the linearized description breaks down. For a black hole one
might formally take the characteristic gravitational scale to be
\(R_S = 2GM\), but then one should not interpret the above linearized calculation
as a controlled derivation of a black-hole condensate. At most, it shows that
the exterior Newtonian dressing has the same parametric mass dependence often
emphasized in corpuscular descriptions \cite{Ruffini:1969qy,Dvali:2011aa,Casadio:2021eio,Casadio:2017cdv,Casadio:2016zpl,Muck:2014kea},
\begin{equation}
{\cal N}_{\rm dress}
\sim
\left(\frac{M}{M_{\rm P}}\right)^2 ,
\end{equation}
modulo the logarithmic dependence on the chosen infrared and ultraviolet
regulators.

This result should be distinguished from the occupation number of physical
gravitons. A static Newton field does not correspond to a collection of on-shell
transverse-traceless quanta. The TT number operator annihilates the unshifted TT
vacuum sector of the state constructed in this paper. Thus any literal statement
such as
\begin{equation}
N_{\rm phys}
=
\int \frac{d^3 k}{(2\pi)^3}
a^{{\rm TT}\,\dagger}_{ij}(k)a^{\rm TT}_{ij}(k)
\end{equation}
would give no coherent TT population for the static field. The nonzero quantity
above is instead a measure of the scalar constraint-sector dressing in the
unreduced BRST description.

This distinction is essential. The coherent state derived in this work is
physical only through its BRST-invariant content and through gauge-invariant
observables such as the Newton potential, curvature components, and boundary
charges. The dressing mode-count \({\cal N}_{\rm dress}\) is therefore best
understood as a useful diagnostic of semiclassicality: large \(GM^2\) means that
the classical Newton field is represented by a large coherent displacement in
the constrained sector. It is not, by itself, a gauge-invariant particle number.

The connection with corpuscular gravity should therefore be stated in the
following limited sense. The BRST coherent-state construction reproduces the
same parametric scaling,
\begin{equation}
{\cal N}_{\rm dress}
\sim
\left(\frac{M}{M_{\rm P}}\right)^2,
\end{equation}
which is often associated with the number of soft gravitational quanta in a
classical field. However, in the present derivation these modes are not
propagating TT gravitons; they are the non-radiative scalar modes required by
the gravitational constraints. The result supports the broader intuition that a
macroscopic gravitational field corresponds to a large quantum displacement, but
it does not by itself establish a gauge-invariant graviton condensate picture.

In summary, the static Newton field carries a large coherent BRST dressing whose
size scales as
\begin{equation}
{\cal N}_{\rm dress}
\sim
G M^2
\log\left(\frac{\Lambda}{\mu}\right).
\end{equation}
This scaling is physically meaningful as a measure of the semiclassical size of
the gravitational dressing, but not as a literal count of physical on-shell
gravitons. The central conclusion of the paper is therefore preserved: the
Newton field is encoded in the scalar constraint sector, while the propagating
transverse-traceless graviton sector remains unexcited.

\section{Conclusion: asymptotic gravity states, holography, and black hole information}

We have derived the quantum state of the asymptotic linearized gravitational field of a compact static source by imposing the sourced BRST condition.  The central result is that the static Newton field is not a coherent state of propagating transverse-traceless gravitons.  It is a coherent state in the non-radiative scalar constraint sector of the unreduced BRST Hilbert space.  The relevant source appears in the Hamiltonian constraint,
\begin{equation}
        {\cal H}_M(x)={\cal H}(x)-16\pi G\rho(x),
\end{equation}
and therefore shifts the scalar null oscillator \(A(\mathbf k)\) appearing in the positive-frequency part of the constraint:
\begin{equation}
        A(\mathbf k)\longrightarrow A(\mathbf k)-q_M(\mathbf k).
\end{equation}
For a regulated compact source of total energy \(E\), the resulting BRST state has the schematic form
\begin{equation}
        |g(E);a\rangle_{\rm BRST}
        =
        \exp\!\left[
        \int \frac{d^3 k}{(2\pi)^3}\,
        q_E(\mathbf k) B^\dagger(\mathbf k)
        \right]
        |0\rangle_{\rm BRST},
\end{equation}
with
\begin{equation}
        q_E(\mathbf k)
        =
        2\sqrt{\pi G}\,
        \frac{E\,\widetilde f_a(\mathbf k)}{k^{3/2}},
\end{equation}
up to the sign and normalization conventions fixed in the main text.  This state satisfies
\begin{equation}
        \langle g(E);a|{\cal H}(\mathbf k)|g(E);a\rangle
        =
        16\pi G\,E\,\widetilde f_a(\mathbf k),
\end{equation}
and hence
\begin{equation}
        \langle \Phi(\mathbf k)\rangle_E
        =
        -\frac{4\pi G E\,\widetilde f_a(\mathbf k)}{k^2}.
\end{equation}
In the point-source limit this becomes the ordinary Newton potential,
\begin{equation}
        \langle \Phi(\mathbf x)\rangle_E
        \longrightarrow
        -\frac{GE}{r}.
\end{equation}

Now consider a compact quantum source prepared in a superposition of energy eigenstates, with some spread in energies concentrated around a semiclassical value:
\begin{equation}
        |\psi\rangle_{\rm src}
        =
        \sum_n c_n |E_n\rangle_{\rm src}.
\end{equation}
The BRST construction associates to each energy eigenvalue \(E_n\) a corresponding asymptotic scalar gravity state,
\begin{equation}
        |E_n\rangle_{\rm src}
        \quad\longmapsto\quad
        |g(E_n);a\rangle_{\rm BRST}.
\end{equation}
Thus the joint source-gravity state is
\begin{equation}
        |\Psi\rangle
        =
        \sum_n c_n
        |E_n\rangle_{\rm src}
        \otimes
        |g(E_n);a\rangle_{\rm BRST}.
\end{equation}
If we focus on the asymptotic gravitational state itself, suppressing the source label, we obtain
\begin{equation}
        |\Psi\rangle_{\rm grav}
        =
        \sum_n c_n |g(E_n);a\rangle_{\rm BRST}.
\end{equation}
This is the gravity analogue of the exterior-geometry state written by Calmet and Hsu \cite{Calmet:2021cip} as
\begin{equation}
        |\Psi\rangle
        =
        \sum_n c_n |g(E_n)\rangle .
\end{equation}

The map from source energies to asymptotic gravity states is already visible at the semiclassical level:
\begin{equation}
        E_n
        \longmapsto
        \langle \Phi(\mathbf x)\rangle_{E_n}
        =
        -GE_n
        \int d^3y\,\frac{f_a(\mathbf y)}{|\mathbf x-\mathbf y|}.
\end{equation}
Equivalently, the ADM energy can be read from the \(1/r\) coefficient of the asymptotic potential:
\begin{equation}
        E_n
        =
        -\frac{1}{G}
        \lim_{r\to\infty}
        r\,\langle \Phi(r)\rangle_{E_n}.
\end{equation}
In operator language, the boundary Hamiltonian is the surface term associated with the gravitational constraint.  At spatial infinity it gives
\begin{equation}
        H_{\partial}
        =
        -\frac{1}{4\pi G}
        \lim_{r\to\infty}
        \int_{S^2_r} d\Omega\, r^2\,\partial_r \Phi,
\end{equation}
so that
\begin{equation}
        H_{\partial}|g(E_n);a\rangle_{\rm BRST}
        =
        E_n |g(E_n);a\rangle_{\rm BRST}.
\end{equation}
Thus the energy decomposition of the compact source is encoded directly in the asymptotic gravity state.

More generally, suppose the distinct asymptotic gravity states
\begin{equation}
        |g_n\rangle \equiv |g(E_n);a\rangle_{\rm BRST}
\end{equation}
are linearly independent.  Define their Gram matrix
\begin{equation}
        G_{mn}
        =
        \langle g_m|g_n\rangle .
\end{equation}
On the subspace spanned by these states, the dual basis is
\begin{equation}
        \langle \widetilde g^{\,n}|
        =
        \sum_m (G^{-1})^{nm}\langle g_m|.
\end{equation}
Then the coefficients of the original source state can be reconstructed from the asymptotic gravity state by
\begin{equation}
        c_n
        =
        \langle \widetilde g^{\,n}|\Psi\rangle_{\rm grav}.
\end{equation}
At leading order in the static Newton field, this reconstruction distinguishes energy eigenvalues.  If there are exact degeneracies among compact-source states with the same \(E_n\), the leading \(1/r\) field alone is insufficient.  However, the full quantum gravitational dressing is not exhausted by the leading Newton potential.  Quantum corrections to the exterior gravitational field, including loop-induced long-distance terms and higher multipole/gravitational dressing data, can depend on the internal quantum state of the compact source.  In that sense the complete asymptotic gravitational state can carry more information than the classical no-hair data.

This conclusion is suggestive of holography \cite{Maldacena:1997re}.  The information about the compact source is not localized purely in the interior matter degrees of freedom.  The gravitational constraints tie the bulk state to asymptotic boundary data.  In the present linearized calculation this statement is concrete: the Hamiltonian constraint fixes the scalar BRST coherent state at infinity, and the coefficient of that state determines the source energy.  In a complete theory of quantum gravity, this is the mechanism by which boundary observables can encode bulk information.  The result is not yet a full derivation of holography, but it exhibits the same structural feature: physical information about the bulk is redundantly available in the gravitational field at the boundary.

The same logic has direct implications for black hole evaporation.  Let the initial black hole state be a superposition of energy eigenstates,
\begin{equation}
        |\psi\rangle_{\rm BH}
        =
        \sum_n c_n |E_n\rangle_{\rm BH}.
\end{equation}
The exterior gravitational field is then
\begin{equation}
        |\Psi\rangle_{\rm ext}
        =
        \sum_n c_n |g(E_n)\rangle ,
\end{equation}
where \(|g(E_n)\rangle\) now denotes the quantum state of the exterior geometry, including the scalar constraint-sector dressing derived in this paper.  The emission amplitude for a Hawking quantum depends on the exterior geometry state.  Following the notation of Calmet and Hsu \cite{Calmet:2021cip}, denote the full set of quantum numbers of the \(i\)-th emitted particle by
\begin{equation}
        r_i\sim\{\omega_i,\mathbf p_i,s_i,q_i,\ldots\},
\end{equation}
and write the amplitude for emission of \(r_i\) from an exterior geometry of energy \(E\) as
\begin{equation}
        \alpha(E,r_i).
\end{equation}
After one emission, the state evolves as
\begin{equation}
        \sum_n c_n |g(E_n)\rangle
        \longrightarrow
        \sum_n\sum_{r_1}
        c_n\,\alpha(E_n,r_1)
        |g(E_n-\omega_1),r_1\rangle .
\end{equation}
After a second emission,
\begin{equation}
        \sum_n\sum_{r_1,r_2}
        c_n\,
        \alpha(E_n,r_1)
        \alpha(E_n-\omega_1,r_2)
        |g(E_n-\omega_1-\omega_2),r_1,r_2\rangle .
\end{equation}
Iterating, the final radiation state takes the form
\begin{align}
        |\Psi\rangle_{\rm rad}
        =
        \sum_n
        \sum_{r_1,\ldots,r_N}
        c_n\,
        &\alpha(E_n,r_1)
        \alpha(E_n-\omega_1,r_2)
        \alpha(E_n-\omega_1-\omega_2,r_3)
        \cdots
        \nonumber\\
        &\times
        |r_1 r_2\cdots r_N\rangle ,
\end{align}
where the final geometry label has been omitted because the black hole has completely evaporated.  The important point is that the amplitudes in this expression are not universal thermal amplitudes defined on a fixed classical background.  They depend on the quantum exterior geometry, and therefore on the branch of the black hole state.

This is the quantum-hair mechanism in the language of the present BRST construction.  The black hole state determines the asymptotic gravitational state.  The asymptotic gravitational state affects the Hawking amplitudes.  Therefore the coefficients \(c_n\) of the initial black hole state can influence the detailed final radiation state.  The final state is a coherent superposition over radiation histories, not a tensor product of an unchanged black hole interior with independent thermal Hawking pairs.  In particular, the factorized nice-slice form \cite{Calmet:2024tyu}
\begin{equation}
        |\psi\rangle_{\rm BH}
        \otimes
        \prod_i
        \left(
        \frac{1}{\sqrt2}|0\rangle_{e_i}|0\rangle_{b_i}
        +
        \frac{1}{\sqrt2}|1\rangle_{e_i}|1\rangle_{b_i}
        \right)
\end{equation}
does not describe the full gravitational Hilbert space.  It freezes the geometry and omits the branch-dependent exterior gravity states \(|g(E_n)\rangle\).

The picture that emerges is conservative but powerful.  The BRST constraints imply that the quantum state of a compact mass is inseparable from its asymptotic gravitational field.  For ordinary compact sources this gives a precise quantum version of the Newton field.  For black holes it implies that the exterior quantum geometry carries information about the black hole state, and that this information can enter the Hawking radiation amplitudes during evaporation.  The calculation in this paper supplies the explicit linearized BRST origin of that exterior state: it is a scalar constraint-sector coherent state, not a cloud of propagating gravitons.  This provides a concrete mechanism by which gravitational constraints, boundary data, and quantum hair are tied together.

\bigskip

\noindent {\it Acknowledgments:}
We used several AI models in the preparation of this manuscript, including GPT-5, DeepSeek v4, DeepMind Co-Scientist, and Gemini DeepThink. The last two models are not available to the general public. The authors have reviewed the manuscript and are responsible for all errors and omissions. We thank Elahe Vedadi, Juraj Gottweis, and Vivek Natarajan of Google/DeepMind for conversations and assistance with AI tools. See \cite{HS} for remarks on the use of generative AI in theoretical physics research.

\noindent {\it Data Availability Statement:}
	This manuscript has no associated data. Data sharing not applicable to this article as no datasets were generated or analysed during the current study.

\appendix

\section{Fourier conventions and useful identities}

For convenience, let us list the Fourier conventions used throughout:
\begin{equation}
f(\mathbf x)=\int \frac{d^3k}{(2\pi)^3}e^{i\mathbf k\cdot\mathbf x}\tilde f(\mathbf k),
\qquad
\tilde f(\mathbf k)=\int d^3x\, e^{-i\mathbf k\cdot\mathbf x}f(\mathbf x).
\end{equation}
Under these conventions,
\begin{equation}
\partial_i f(\mathbf x)\longleftrightarrow ik_i\tilde f(\mathbf k),
\qquad
\nabla^2 f(\mathbf x)\longleftrightarrow -k^2\tilde f(\mathbf k).
\end{equation}
The Green's-function identity used repeatedly in the paper is
\begin{equation}
\int \frac{d^3k}{(2\pi)^3}\,\frac{e^{i\mathbf k\cdot\mathbf r}}{k^2}=\frac{1}{4\pi |\mathbf r|}.
\end{equation}
Equivalently,
\begin{equation}
\nabla^2\left(-\frac{1}{r}\right)=4\pi\delta^{(3)}(\mathbf r).
\end{equation}

\section[Explicit check of the coherent-state expectation value]{Explicit check of the coherent-state\\ expectation value}

It is useful to check directly that the state \eqref{eq:gravitycoherentstate} gives the expected one-point function of $A$. Write
\begin{equation}
\ket{M;a}_S=e^X\ket{0}_S,
\qquad
X=\int \frac{d^3k}{(2\pi)^3}\,q_M(\mathbf k)B^\dagger(\mathbf k).
\end{equation}
Then
\begin{equation}
A(\mathbf p)e^X=e^X(A(\mathbf p)+q_M(\mathbf p)).
\end{equation}
So
\begin{equation}
A(\mathbf p)\ket{M;a}_S=q_M(\mathbf p)\ket{M;a}_S.
\end{equation}
The state is therefore an eigenstate of $A(\mathbf p)$ with eigenvalue $q_M(\mathbf p)$. Taking the Hermitian adjoint and using real $q_M$, we similarly get
\begin{equation}
{}_S\bra{M;a}A^\dagger(\mathbf p)={}_S\bra{M;a}q_M(\mathbf p).
\end{equation}
Hence
\begin{equation}
{}_S\bra{M;a}\mathcal H(\mathbf k)\ket{M;a}_S
=\frac{\kappa k^{3/2}}{\sqrt2}\,\big(q_M(\mathbf k)+q_M(\mathbf k)\big)
=\sqrt2\,\kappa k^{3/2}q_M(\mathbf k),
\end{equation}
as used in the main text.

\section{Deriving the form of \texorpdfstring{$B^\dagger(\mathbf k)$}{B dagger(k)} in the coherent state}

The construction in the main text gives an explicit canonical form for the null partner:
\begin{equation}
B(\mathbf k)=-\left(\frac{1}{4}\theta^{ij}+\ell^{ij}\right)a_{ij}(\mathbf k),
\qquad
B^\dagger(\mathbf k)=-\left(\frac{1}{4}\theta^{ij}+\ell^{ij}\right)a^\dagger_{ij}(\mathbf k),
\label{eq:BdaggerCanonicalAppendix}
\end{equation}
where
\begin{equation}
\theta_{ij}=\delta_{ij}-\hat k_i\hat k_j,
\qquad
\ell_{ij}=\hat k_i\hat k_j.
\end{equation}
This is the operator that appears in
\begin{equation}
\ket{M;a}_S
=
\exp\!\left[\int \frac{d^3k}{(2\pi)^3}\,q_M(\mathbf k)B^\dagger(\mathbf k)\right]\ket{0}_S.
\end{equation}

Let us verify the two required commutators directly. From the spatial graviton oscillator algebra,
\begin{equation}
[s^{ij}a_{ij}(\mathbf k),t^{\ell m}a^\dagger_{\ell m}(\mathbf k')]
=(2\pi)^3\delta^{(3)}(\mathbf k-\mathbf k')
\left(s^{ij}t_{ij}-\frac{1}{2}s^i{}_i t^j{}_j\right).
\end{equation}
The active constrained mode is
\begin{equation}
A(\mathbf k)=\theta^{ij}a_{ij}(\mathbf k).
\end{equation}
Using
\begin{equation}
\langle\theta,\theta\rangle_{\rm DW}=0,
\qquad
\langle\theta,\ell\rangle_{\rm DW}=-1,
\qquad
\langle\ell,\ell\rangle_{\rm DW}=\frac{1}{2},
\end{equation}
one finds
\begin{align}
[A(\mathbf k),B^\dagger(\mathbf k')]
&=-(2\pi)^3\delta^{(3)}(\mathbf k-\mathbf k')
\left\langle\theta,\frac{1}{4}\theta+\ell\right\rangle_{\rm DW} \\
&=(2\pi)^3\delta^{(3)}(\mathbf k-\mathbf k'),
\end{align}
and
\begin{align}
[B(\mathbf k),B^\dagger(\mathbf k')]
&=(2\pi)^3\delta^{(3)}(\mathbf k-\mathbf k')
\left\langle \frac{1}{4}\theta+\ell,\frac{1}{4}\theta+\ell\right\rangle_{\rm DW} \\
&=0.
\end{align}
Thus $B^\dagger$ is the null creation operator conjugate to the active constrained annihilator $A$.

Finally, define
\begin{equation}
X:=\int \frac{d^3k}{(2\pi)^3}\,q_M(\mathbf k)B^\dagger(\mathbf k).
\end{equation}
Then
\begin{align}
[A(\mathbf p),X]
&=\int \frac{d^3k}{(2\pi)^3}\,q_M(\mathbf k)[A(\mathbf p),B^\dagger(\mathbf k)] \\
&=q_M(\mathbf p),
\end{align}
so all higher nested commutators vanish and
\begin{equation}
(A(\mathbf p)-q_M(\mathbf p))e^X=e^X A(\mathbf p).
\end{equation}
Acting on the scalar vacuum gives
\begin{equation}
(A(\mathbf p)-q_M(\mathbf p))\ket{M;a}_S=0.
\end{equation}

The choice of $B^\dagger$ is not unique as a representative in the unreduced indefinite-metric state space. Any replacement
\begin{equation}
B^\dagger\longrightarrow B^\dagger+Y^\dagger,
\qquad
[A,Y^\dagger]=0,
\end{equation}
produces the same displacement of $A$. If one also demands that the new partner remain a null oscillator, then $Y^\dagger$ must obey the corresponding additional null condition. The source fixes the shifted constrained mode and hence the Newton field; it does not fix a unique unreduced Fock-space dressing.


\end{document}